# Lattice parameter, lattice disorder and resistivity of carbohydrate doped MgB$_2$ and their correlation with the transition temperature

J.H. Kim [1, *], Sangjun Oh [2], X. Xu [1], Jinho Joo [3], M. Rindfleisch [4], M. Tomsic [4], S.X. Dou [1]

[1] *Institute for Superconducting and Electronic Materials, University of Wollongong, Northfields Avenue, Wollongong, NSW 2522, Australia*
[2] *Material Research Team, National Fusion Research Institute, 52 Eoeun-dong, Yuseong, Daejeon, Chungnam 305-333, Korea*
[3] *School of Advanced Material Science and Engineering, Sungkyunkwan Unversity, Suwon, Gyeonggi 440-746, Korea*
[4] *Hyper Tech Research, Incorporated, 1275 Kinnear Road, Columbus, OH 43212, USA*

The change in the lattice parameters or the lattice disorder is claimed as a cause of the slight reduction in the transition temperature by carbon doping in MgB$_2$. In this work, an extensive investigation on the effects of carbohydrate doping has been carried out. It is found that not only the *a*-axis but also the *c*-axis lattice parameter increases with the sintering temperature. A linear relation between the unit cell volume and the critical temperature is observed. Compared with the well known correlation between the lattice strain and the critical temperature, the X-ray peak broadening itself shows a closer correlation with the transition temperature. The residual resistivity and the critical temperature are linearly correlated with each other as well and its implication is further discussed.

## 1. INTRODUCTION

It is well known that carbon substitutes born at the boron sites of MgB$_2$ results in decrease of the lattice parameters [1]. There can be a decrease in the density of state (DOS) and hardening of the optical $E_{2g}$ phonons which is known to be strongly related with the superconducting properties of MgB$_2$ [2]. It was further reported that doping induced disorder strongly affects the superconducting properties of MgB$_2$. In early works of Serquis *et. al.*, it was shown that lattice strain is strongly related with resistivity and the transition temperature as well [3,4]. In a recent review article of Eisterer, an interesting correlation between the normalized resistivity, defined as a ratio of the resistivity at 40 K to the difference in resistivity at 300 and 40 K, and the transition temperature was argued from the analysis of various data reported [1]. Also, it was claimed that even the upper critical field can be represented as a function of the transition temperature and the normalized resistivity. We recently reported on the effects of carbohydrate doping on the superconducting properties of MgB$_2$ and the importance of lattice disorder was argued [5]. The study of carbon doping using carbohydrate can be beneficial since highly uniform mixing is possible compared with other nano carbon doping methods. In this work, as an extension of our previous work on carbohydrate doping, the variation of various parameters related with the lattice constant with the change in sintering temperature has been extensively studied. The change in the lattice parameter and resistivity, and their correlation with the transition temperature will be discussed.

## 2. EXPERIMENTAL DETAILS

a) Author to whom correspondence should be address.
Electronic mail: jhk@uow.edu.au

Carbohydrate doped MgB$_2$ wire was fabricated by an *in situ* powder-in-tube (PIT) process. MgB$_2$ + 10wt% C$_4$H$_6$O$_5$ powder was prepared by using a chemical solution route as was reported in our previous work [5]. The powder was packed into a 140 mm long iron (Fe) tube. The outer diameter of the Fe tube was 10 mm and the inner diameter was 8 mm. The packed tube was drawn till the final outer diameter became 1.42 mm. The fabricated wires were sintered at a wide range of temperatures from 650 to 1000 $^o$C for 30 min under high purity argon gas with a ramp rate of 5 $^o$Cmin$^{-1}$.

All samples were characterized by detailed X-ray diffraction analysis and by resistivity measurement. PW1730 X-ray diffractometer was used for the determination of the phase and crystal structure. The *a*- and *c*-lattice parameters were obtained from the Rietveld refinement using Fullprof software. The grain size and lattice strain values were estimated from Williamson-Hall plots. Transport measurements for resistivity were carried out by the standard ac four-probe method.

## 3. RESULTS AND DISCUSSION

The changes in the lattice parameters with the variation of the sintering temperature are presented in Figure 1. As was reported in our previous work [5], the *a*-axis of the carbohydrate doping sample slightly increases from 3.0757 Å to 3.0780 Å as the sintering temperature is raised. This trend is rather peculiar compared with the reported features for the other solid state carbon doping methods. Usually, the opposite behavior is observed even for the nano SiC doping [6]. From the comparison with single crystal results, the amount of carbon substitution can be inferred from the lattice parameter [7]. Less carbon is incorporated into the boron site as the sintering temperature increased for the carbohydrate doping [5]. The more weird result is that the *c*-axis lattice parameter is also increased with the sintering temperature as can be seen in Figure 1(b). For the other carbon doping cases, the *c*-axis lattice parameter is reported to be more or less the same with the sintering temperature. The relative change in the *c*-axis is almost 3.2 times larger more than that of the *a*-axis lattice parameter and the *c/a* ratio mimics the *c*-axis variation (Figure 1(c)). The carbohydrate doping leads to an anisotropic contraction of the unit cell (Figure 1(d)), which might result from the difference in the bond strength [8]. The transition temperature also gradually increases with the sintering temperature, as can be seen in Figure 2. We can observe a correlation between the lattice parameter and the transition temperature as shown in Figure 3. In particular, a strong linear correlation is observed between the unit cell volume and the transition temperature with a slope of 1 K/0.0983 Å$^3$ comparable with previous reports [9].

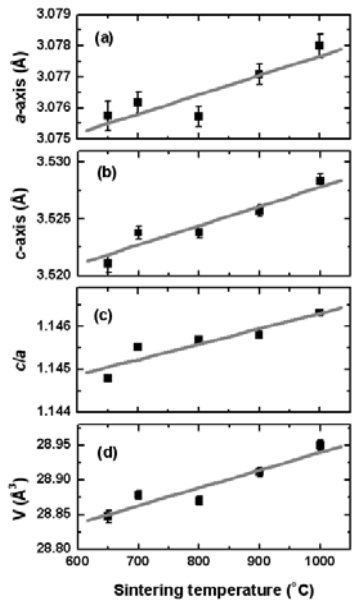

**Fig. 1.** (a) *a*-axis lattice parameter, (b) *c*-axis lattice parameter, (c) *c/a* values, and (d) unit cell volume of $MgB_2$ + 10wt% $C_4H_6O_5$ wires calculated from Rietveld refinement. All lines are guides to the eyes.

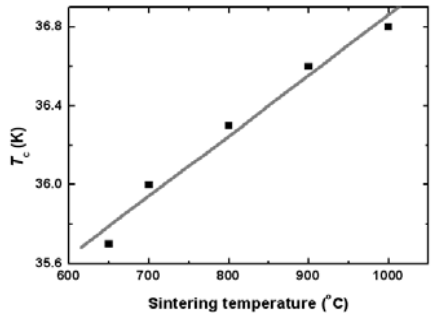

**Fig. 2.** Transition temperature behavior ($T_c$) of $MgB_2$ + 10wt% $C_4H_6O_5$ wires as a function of sintering temperature. Line is guides to the eyes.

For nano SiC doping, the similar trend is observed in the transition temperature variation with the sintering temperature [6]. However, the transition temperature of the carbohydrate sample is slightly higher than that of the nano SiC doped sample. At the sintering temperature of 1000 °C, the transition temperature is as high as ~36.8 K. The difference in the transition temperature observed among various carbon doped samples might be also related with the impurity scattering between $\sigma$ and $\pi$ bands of $MgB_2$ caused by lattice disorder [1]. The correlation between the lattice strain and the transition temperature was first reported by Serquis *et. al.* [3] and we also found that such a correlation exists for our carbohydrate samples [5]. The lattice strain, $\varepsilon$ and the average grain size, $L$ can be estimated from a Williamson-Hall plot for the X-ray diffraction peak broadenings [10],

$$\beta \cos\theta = 4\varepsilon \sin\theta + \lambda / L \qquad (1)$$

where, $\beta$ is the full width at half maximum (FWHM) in radian and $\lambda$ is the wavelength. A closer correlation can be found between the FWHM and the transition temperature compared with the correlation between the lattice stain and the critical temperature. Improved crystallinity due to the grain growth might reduce defects and the overall peak broadening can be a better representative for the lattice disorder. The grain size also linearly increased with the sintering temperature as shown in Figure 4. We already reported that the pinning force is actually decreased as the sintering temperature raised. Our result is consistent with the idea that the grain boundary pinning is the major pining mechanism for $MgB_2$ [1].

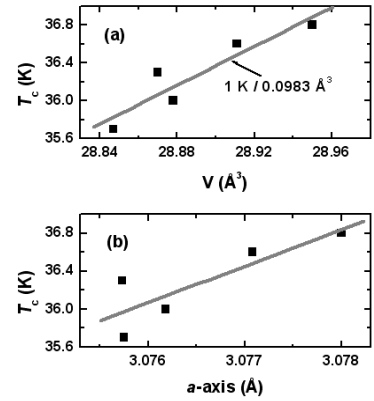

**Fig. 3.** Influence of (a) unit cell volume and (b) *a*-axis lattice parameter on the transition temperature ($T_c$). All lines are guides to the eyes.

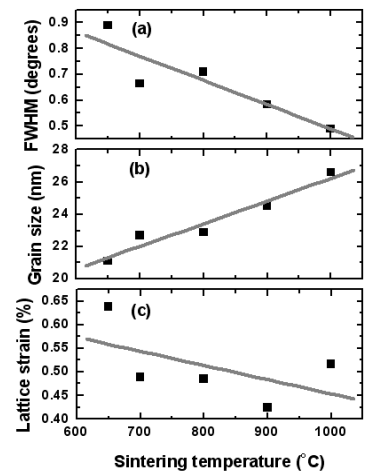

**Fig. 4.** (a) Full width at half maximum, (b) grain size, and (c) micro strain as a function of sintering temperature ($T_c$). All lines are guides to the eyes.

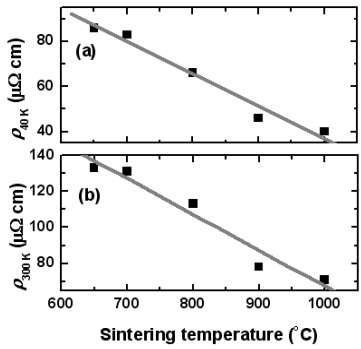

**Fig. 5.** (a) $\rho_{40\,K}$ and (b) $\rho_{300\,K}$ as a function of sintering temperature. All lines are guides to the eyes.

Lattice disorder also can influence the resistivity ($\rho$). Both the residual resistivity, $\rho_{40K}$ and room temperature resistivity, $\rho_{300K}$ were linearly decreased with the sintering temperature as shown in Figure. 5. We reported that there is a linear relation between the normalized resistivity, which is defined as a ratio between the residual resistivity and the difference of the room temperature resistivity from the residual resistivity ($\rho_{norm} = \rho_{40K}/\Delta\rho$, $\Delta\rho = \rho_{300K} - \rho_{40K}$), and the transition temperature [5]. Similar linear relations can be found between the residual resistivity and the transition temperature, between the room temperature resistivity and the transition temperature, as well. On the other hand, no such a correlation is observed between $\Delta\rho$ and the critical temperature as can be seen in Figure. 6. It is generally argued that the residual resistivity, $\rho_{40K}$ is related with the intragrain impurity scattering whereas the difference between the residual resistivity and room temperature resistivity, $\Delta\rho = \rho_{300K} - \rho_{40K}$ is affected by the intergrain scattering [1,11]. As was noted, it was argued that the impurity scattering between $\sigma$ and $\pi$ bands can affect the critical temperature of $MgB_2$ [1]. The intragrain impurity scattering seems to be closely correlated with the impurity scattering between $\sigma$ and $\pi$ bands results in the close correlation with the critical temperature.

## 4. CONCLUSIONS

Detailed X-ray diffraction analysis has been carried out for the carbohydrate doped $MgB_2$ with the variation in the sintering temperature. Especially, we found that the *c*-axis lattice parameter is increased with the sintering temperature contrary to previous studies on solid state carbon doped samples. By carbohydrate doping, the unit cell is anisotropically contracted and there is a linear relation between the unit cell volume and the transition temperature with a slope of 1 K/0.0983 Å$^3$. Lattice disorder was estimated from the X-ray diffraction peak broadenings and we found a close correlation between the peak broadening itself and the critical temperature. Furthermore, a linear relation between the residual resistivity and the critical temperature is observed whereas no correlation is found between $\Delta\rho$ and the transition temperature, which seems to be related with the doing induced impurity scattering between $\sigma$ and $\pi$ bands of $MgB_2$ within grain.

**Acknowledgements:** This work was supported by the Australian Research Council, Hyper Tech Research Inc., OH, USA, and Alphatech International Ltd, NZ. The work done at the National Fusion Research Institute was supported by a Korea Science and Engineering Foundation (KOSEF) grant funded by the Korean Government (MEST) (No. R01-2007-000-20462-0).

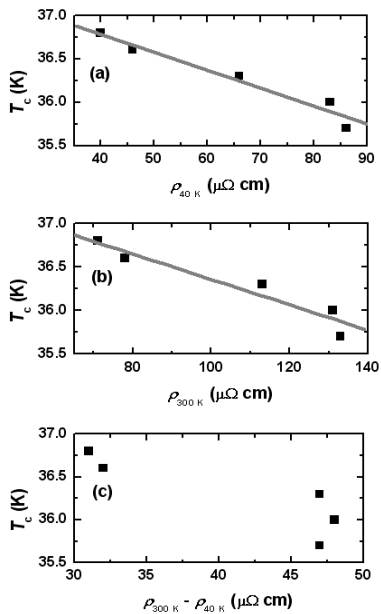

**Fig. 6.** Influence of (a) $\rho_{40\,K}$, (b) $\rho_{300\,K}$, and (c) $\Delta\rho = \rho_{300\,K} - \rho_{40\,K}$ on the transition temperature ($T_c$). All lines are guides to the eyes.